\documentclass[twoside]{article}
\usepackage{iwce}
\usepackage{amsmath,amsfonts,bm,graphicx}
\begin{document}

\title{Modelling of Quantum Electromechanical Systems}

\IWCEauthorsFirst{ANTTI-PEKKA JAUHO, TOM\'A\v S NOVOTN\'Y, ANDREA
DONARINI, and CHRISTIAN FLINDT} \setHeadings{Jauho}{Modelling of
Quantum Electromechanical Systems} \IWCEaddressFirst{MIC --
Department of Micro and
Nanotechnology, Technical University of Denmark, Bldg. 345East\\
DK-2800 Lyngby, Denmark} \email{antti@mic.dtu.dk}


\preparetitle

\begin{IWCEabstract}
We discuss methods for numerically  solving the generalized Master
equation GME which governs the time-evolution of the reduced
density matrix $\hat\rho$ of a mechanically movable mesoscopic
device in a dissipative environment. As a specific example, we
consider the quantum shuttle -- a generic quantum
nanoelectromechanical system (NEMS). When expressed in the
oscillator basis, the static limit of the GME becomes a large
linear non-sparse matrix problem (characteristic size larger than
$10^4\times 10^4$) which however, as we show, can be treated using
the Arnoldi iteration scheme. The numerical results are
interpreted with the help of Wigner functions, and we compute the
current and the noise in a few representative cases.
\end{IWCEabstract}

\IWCEkeywords{SET, Coulomb blockade, Nanoelectromechanics, Noise}

\begin{multicols}{2}
\intro{Introduction} Microelectromechanical systems (MEMS) are
approaching the nanoscale, which ultimately implies that the
mechanical motion needs to be treated quantum mechanically.  An
example is the experiment by Park et al. \cite{par-nat-00}, where
the break junction technique was used to create a single-electron
transistor with a C$_{60}$-molecule as the active part. The
measured IV-curves display features that can be related to the
mechanical vibrations of the molecule. Already earlier it was
suggested theoretically \cite{gor-prl-98} that a nanoscopic
movable metallic grain, when operated in the Coulomb blockade
regime, can move charges one-by-one between source and drain
contacts.  This orderly transport was coined the shuttle regime.
In recent years our group has developed theoretical methods to
analyze the shuttle transition in the quantum regime
\cite{nov-prl-03,nov-prl-04,fli-prb-04,fli-epl-04}, focusing not
only on the IV-curve, but also considering noise and full counting
statistics, which are important diagnostic tools in unravelling
the microscopic transport mechanisms.  In this paper we present
examples of our numerical results and discuss pertinent
computational issues.

\section{The generalized Master Equation (GME)}
\noindent The Hamiltonian for the system includes terms describing
(i) the electronic part of the movable quantum dot (QD for short),
(ii) its mechanical motion (which is quantized), (iii) the
position dependent coupling of the QD and the leads, (iv) the
leads (treated as noninteracting fermions), and (v) coupling to
environment, which damps the mechanical motion
\cite{nov-prl-03,nov-prl-04,fli-prb-04,fli-epl-04}.  Using methods
familiar from quantum optics, we integrate out the environmental
degrees of freedom (the lead electrons, and a generic heat bath)
to obtain a generalized Master equation for the ``system" (= QD +
quantized oscillator) density operator:
\begin{equation}\dot{\rho}(t)={\mathcal L}\rho(t)
 = (\mathcal{L}_{\rm coh} + \mathcal{L}_{\rm driv}
   + \mathcal{L}_{\rm damp})\rho(t). \label{supermatrix}
\end{equation}
Here $\mathcal{L}_{\rm coh}, \mathcal{L}_{\rm driv}$ and
$\mathcal{L}_{\rm damp}$ are superoperators corresponding to the
coherent evolution, coupling to leads, and damping of the QD. It
is sufficient to consider the diagonal electronic components
(i.e., an empty and an occupied QD, respectively), which satisfy
\cite{nov-prl-03,remark}
\begin{eqnarray}\label{GME}
\dot{\rho}_{00}(t) &=& \frac{1}{i\hbar} [H_{\rm osc},\rho_{00}(t)]
    - \frac{\Gamma_L}{2}(e^{-\frac{2x}{\lambda}}\rho_{00}(t)
    \nonumber\\
&\quad&    + \rho_{00}(t)e^{-\frac{2x}{\lambda}}) + \Gamma_R
e^{\frac{x}{\lambda}}\rho_{11}(t)e^{\frac{x}{\lambda}}
    \nonumber\\
&\quad&    + \mathcal{L}_{\rm damp}\,\rho_{00}(t)\ , \nonumber\\
\dot{\rho}_{11}(t) &=& \frac{1}{i\hbar}[H_{\rm
osc}-eEx,\rho_{11}(t)]
    + \Gamma_L e^{-\frac{x}{\lambda}}\rho_{00}(t)e^{-\frac{x}{\lambda}}
    \nonumber\\
&\quad&    -
    \frac{\Gamma_R}{2}(e^{\frac{2x}{\lambda}}\rho_{11}(t)
     + \rho_{11}(t)e^{\frac{2x}{\lambda}})\nonumber\\
&\quad&    + \mathcal{L}_{\rm damp}\,\rho_{11}(t).
\end{eqnarray}
where
\begin{equation}
   \mathcal{L}_{\rm damp}\rho = -\frac{i\gamma}{2\hbar}[x,\{p,\rho\}] -
   \frac{\gamma m \omega}{\hbar}(\bar{N}+1/2)[x,[x,\rho]]\nonumber \ .
\end{equation}
The physical parameters defining the quantum shuttle are thus the
couplings to leads $\Gamma_{L/R}$, the oscillator frequency
$\omega$, the damping rate of the oscillator $\gamma$, the
temperature $T$, and the tunnelling length $\lambda$.

\section{The numerical solution}
\noindent We seek the steady state,
$\dot\rho_{00}=\dot\rho_{11}=0$, i.e., the null vector of the
Liouvillean, $\mathcal L \rho=0$.  Suppose we keep $N$ lowest
energy states of the oscillator.  Since $\hat\rho_{00/11}$ are
full matrices in the oscillator basis, they both have $N^2$
elements.  The matrix representation of the Liouville
superoperator has thus $2N^2\times 2N^2$ elements. (The situation
is even more demanding if one considers  a shuttle with more than
just two electronic states, see \cite{fli-prb-04,arm-prb-02}.)  We
attempted the solution of these equations with standard Matlab
routines, such as the singular value decomposition, by gradually
increasing the cut-off $N$. As illustrated in  Fig. \ref{fig1},
the Liouville matrices are not sparse. Unfortunately, reliable
convergence could not be achieved once $N$ approached few tens,
which was quite inadequate for physical reasons, which indicated
the necessity of using $N\simeq 100\cdots 200$.
\begin{IWCEfigure}
  \centering
  \includegraphics*[width=\textwidth]{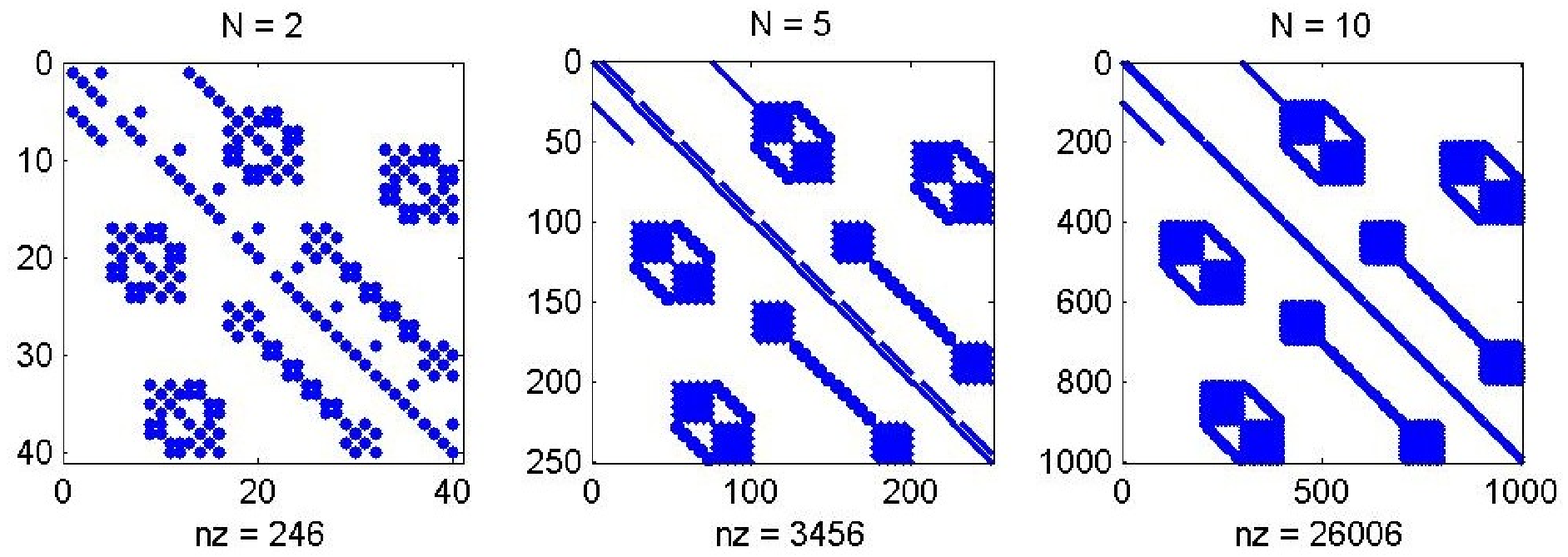}
  \caption{The nonzero matrix elements of the Liouville
  superoperator for a three QD device studied in
  \cite{arm-prb-02,fli-prb-04}
  for three values of the cut-off $N=2,5,10$. $nz$ is the number of
  non-zero elements.}
  \label{fig1}
\end{IWCEfigure}
Clearly, more powerful numerical schemes are required.  One such
method is the Arnoldi iteration \cite{gol-book-96}, which has
clear advantages compared to singular value decomposition both in
terms of computational speed and memory requirements.  First, it
is not necessary to store the full Liouvillean matrix, and,
second, when one seeks  the best approximation to the null vector
it is possible to work in spaces which are much smaller than the
full Liouville space.  The central concept in the Arnoldi scheme
is the Krylov space, defined as
\begin{equation}
\mathcal K_j={\rm span}({\mathbf x}_0,{\mathbf L}{\mathbf
x}_0,...,{\mathbf L}^{j-1}{\mathbf x}_0),
\end{equation}
where $j$ is a {\it small} integer.  The vector ${\mathbf x}_0$ is
a vectorization of some arbitrary state represented by the two
matrices $\hat\rho_{00/11}$, i.e. a vector of length $2N^2$. Note
that the calculation of the vectors ${\mathbf L}^{i}{\mathbf x}_0,
i=1..(j-1)$ can be done directly from the GME, without storing the
full Liouvillean matrix.  The next step consists of finding an
orthonormal basis in the Krylov space, let us denote this by
$\{{\mathbf q}_i\} , i=1,...,j$.  The method proceeds now by
finding an approximate null vector in the Krylov space, spanned by
the vectors $\{{\mathbf q}_i\}$ thus involving matrices of size
$j\times j$. In practice, we have found that $j=20$ is sufficient,
which makes the calculations very fast, and no supercomputing is
necessary. Once the optimal vector in the Krylov space has been
found, we can use the coordinates of this vector to construct the
estimate ${\mathbf x}_1$ (which is a vector of length $2N^2$) from
which the matrices $\rho_{00/11}$ can be reassembled.  If the
result is not sufficiently close to the null vector of the
Liouvillean, one repeats the procedure by constructing the next
Krylov space, now using ${\mathbf x}_1$ as the seed.

The Arnoldi scheme is iterative in its nature, and one must
address the issue of convergence.  For example, it is not a priori
clear how many iterations are needed, and indeed convergence is
not always achieved.  This problem is solved by the use of a
preconditioner.  The basic idea is to find an invertible operator
$\mathcal M$ in the Liouville space, such that the original
problem $\mathcal L \rho^{\rm stat}=0$ is cast in the form
$\mathcal M[\mathcal L[\rho^{\rm stat}]]=0$, and that the
truncated version of the operator $\mathcal M\mathcal L$ gives
rise to a rapidly converging iteration scheme.  The Arnoldi scheme
is particularly efficient in finding good approximations to
eigenvalues (and corresponding eigenvectors) that are separated
from the rest of the spectrum.  Thus, in our case, the
preconditioner should move the eigenvalues with a non-vanishing
real part away from the origin of the complex plane.  A good
candidate for the preconditioner is to use the Sylvester part
\cite{Timo} of $\mathcal L$:
\begin{equation}
\mathcal L[\rho]=\mathcal L_{\rm Sylv}[\rho]+\mathcal L_{\rm
rest}[\rho],
\end{equation}
where the Sylvester part has the structure
\begin{eqnarray}
\mathcal L_{\rm Sylv}[\rho] &=& {\mathbf A}\rho+\rho{\mathbf A}^\dagger\nonumber\\
&=& \begin{pmatrix} A_{00}\rho_{00}+\rho_{00}A_{00}^\dagger & 0 \\
0 & A_{11}\rho_{11}+\rho_{11}A_{11}^\dagger\nonumber
\end{pmatrix},
\end{eqnarray}
where the elements $A_{00/11}$ can be gleaned off from the GME.
The Sylvester part is rapidly invertible, and we thus chose
$\mathcal M=\mathcal L_{\rm Sylv}^{-1}$.  In general, it was
essential to use the preconditioner, however we stress that the
choice is somewhat subtle and there is no unique algorithm for
this.  We have encountered special situations where the Sylvester
preconditioner did not work satisfactorily, and more work is
required in refining the numerics in this case.

Once the static density matrix is solved, the current is readily
calculable from
\begin{eqnarray}
I^{\rm stat}&=&e{\rm Tr}_{\rm osc}\{\Gamma_R
e^{2x/\lambda}\rho^{\rm stat}_{11}\}\nonumber\\
&=&e{\rm Tr}_{\rm osc}\{\Gamma_L e^{-2x/\lambda}\rho^{\rm
stat}_{00}\}.
\end{eqnarray}
We have recently shown \cite{nov-prl-04} that also the noise, and
even the higher cumulants \cite{fli-epl-04}, can be calculated
with similar methods. In particular, we find that the Fano factor
$F=S(0)/2eI$ (here $S(0)$ is the zero-frequency component of the
noise spectrum) can be expressed as
\begin{eqnarray}\label{Fano}
    F &=& 1 - \frac{2e\Gamma_R}{I} \mathrm{Tr_{osc}}\left\{ e^{\frac{2x}{\lambda}}
    \left[\mathcal{Q}\mathcal{L}^{-1}\mathcal{Q} \right.\right.\nonumber\\
&\quad&\times \left.\left.    \begin{pmatrix}
    \Gamma_R e^{\frac{x}{\lambda}}\rho_{11}^{\rm
    stat}e^{\frac{x}{\lambda}}\\
    0 \end{pmatrix}\right]_{11}\right\}\ .
\end{eqnarray}
Here $\mathcal{Q}$ is a projection operator that projects away
from the stationary state. Very importantly, the pseudoinverse
$\mathcal{R}$ of the Liouvillean, defined as
$\mathcal{Q}\mathcal{L}^{-1}\mathcal{Q}\equiv\mathcal{R}$ is
tractable by similar numerical methods as used in the evaluation
of the current (we use the generalized minimum residual method
(GMRes) \cite{fli-msc-04}).  Before showing results for the
current and noise, we discuss an important visualization tool.

\section{Wigner functions}
\noindent We have found that Wigner functions are an excellent
interpretative tool for the numerical results obtained for the
stationary density matrix.  The intuitive picture comes from the
well-known results in the classical limit: the Wigner
representation (or, equivalently, the phase-space representation)
of a regularly moving harmonic oscillator is a circle.  On the
other hand, irregular motion under the influence of external noise
gives rise to a Gaussian probability distribution centered at the
origin.  Since the QD can be either empty or occupied, it is
advantageous to introduce {\it charge resolved} Wigner functions
($n=0$ corresponds to an empty dot, while $n=1$ represents the
occupied dot), defined as
\begin{equation}
    W_{nn}(X,P) =
    \int_{-\infty}^{\infty}\frac{dy}{2\pi\hbar}\,\bigl\langle
    X-\frac{y}{2}\,|\rho_{nn}^{\rm
    stat}|X+\frac{y}{2}\bigr\rangle
    \exp\bigl(i\frac{Py}{\hbar}\bigr).
\end{equation}
An example of the empty dot Wigner function is given in Fig.
\ref{fig2}. The shape is consistent with physical intuition.
Consider, for example, positive $x$ and positive $p$ (the shuttle
is approaching the drain contact).  Then the probability of an
empty dot is large, because the extra charge is very likely to
leave the shuttle as the drain contact is approached. Analogously,
at negative $x$ and positive $p$ the probability of an empty dot
is very small, because the dot has been recharged in the vicinity
of the left (source) contact, and the charge cannot have left the
QD because of the exponentially small coupling to the right
(drain) contact. Very interestingly, in Fig. 2 we also see an
enhanced probability located at the origin of the phase space. The
interpretation is that at these values of parameters there are two
co-existing transport regimes: (i) the charge shuttling regime
(represented by the ring), and (ii) an incoherent tunnelling
regime in which charges tunnel into and out from the QD
uncorrelated to its position.
\begin{IWCEfigure}
  \centering
  \includegraphics*[width=0.8\textwidth]{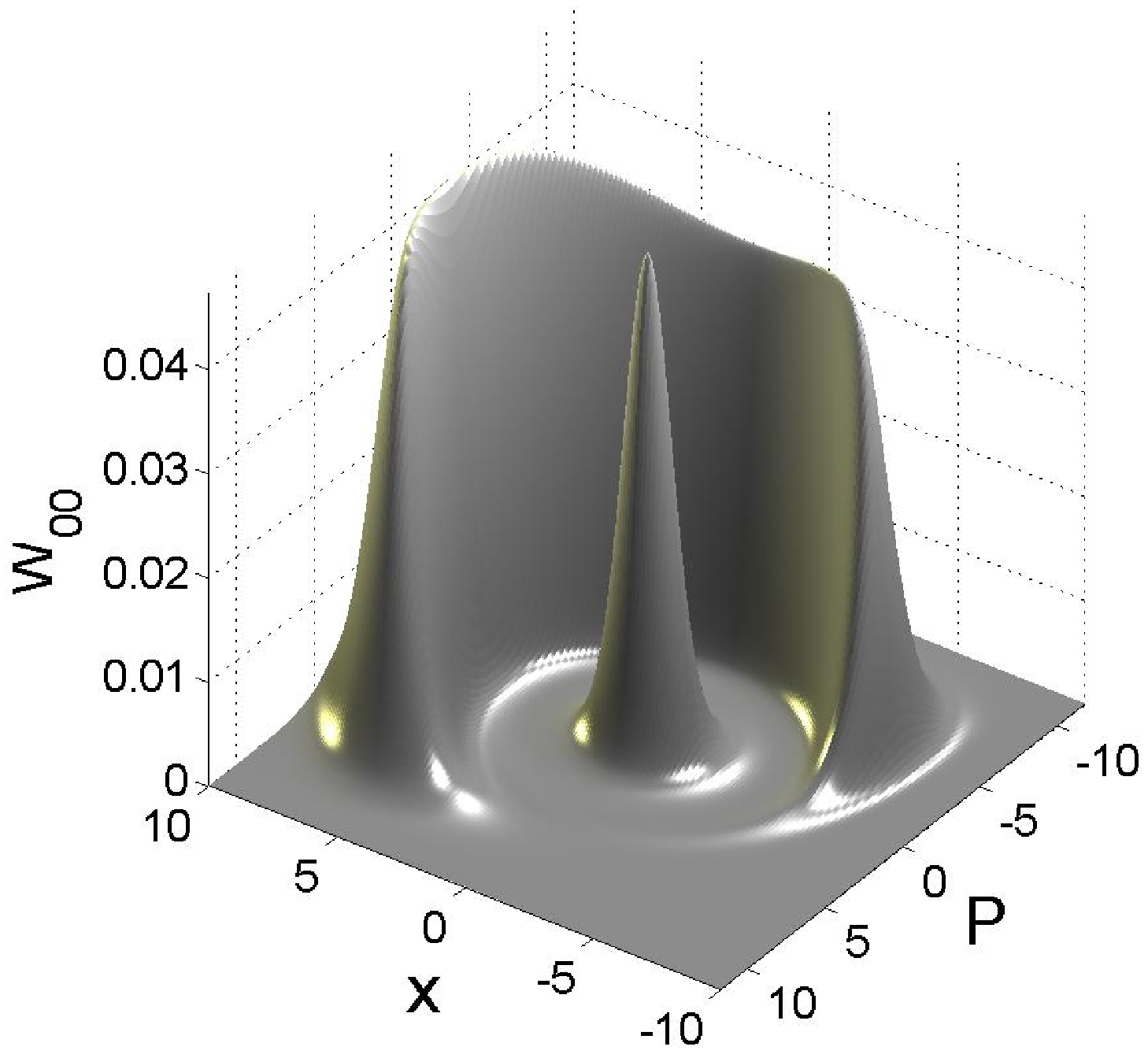}
  \caption{The Wigner representation of the empty-dot density matrix
  in the co-existence regime.}
  \label{fig2}
\end{IWCEfigure}

\section{Numerical results}
\noindent Figure 3 shows the computed stationary current as a
function of the damping rate.  Noteworthy features are: (i) One
observes a shuttling transition (or, perhaps more appropriately, a
cross-over) from a low value of current (tunnelling regime) to a
high value of current (shuttling regime) even in the quantum
regime (small $\lambda$); (ii) in the low damping limit the
current saturates to a universal value of $1/2\pi$ corresponding
to precisely one electron transmitted per cycle.

\begin{IWCEfigure}
  \centering
  \includegraphics*[width=0.8\textwidth]{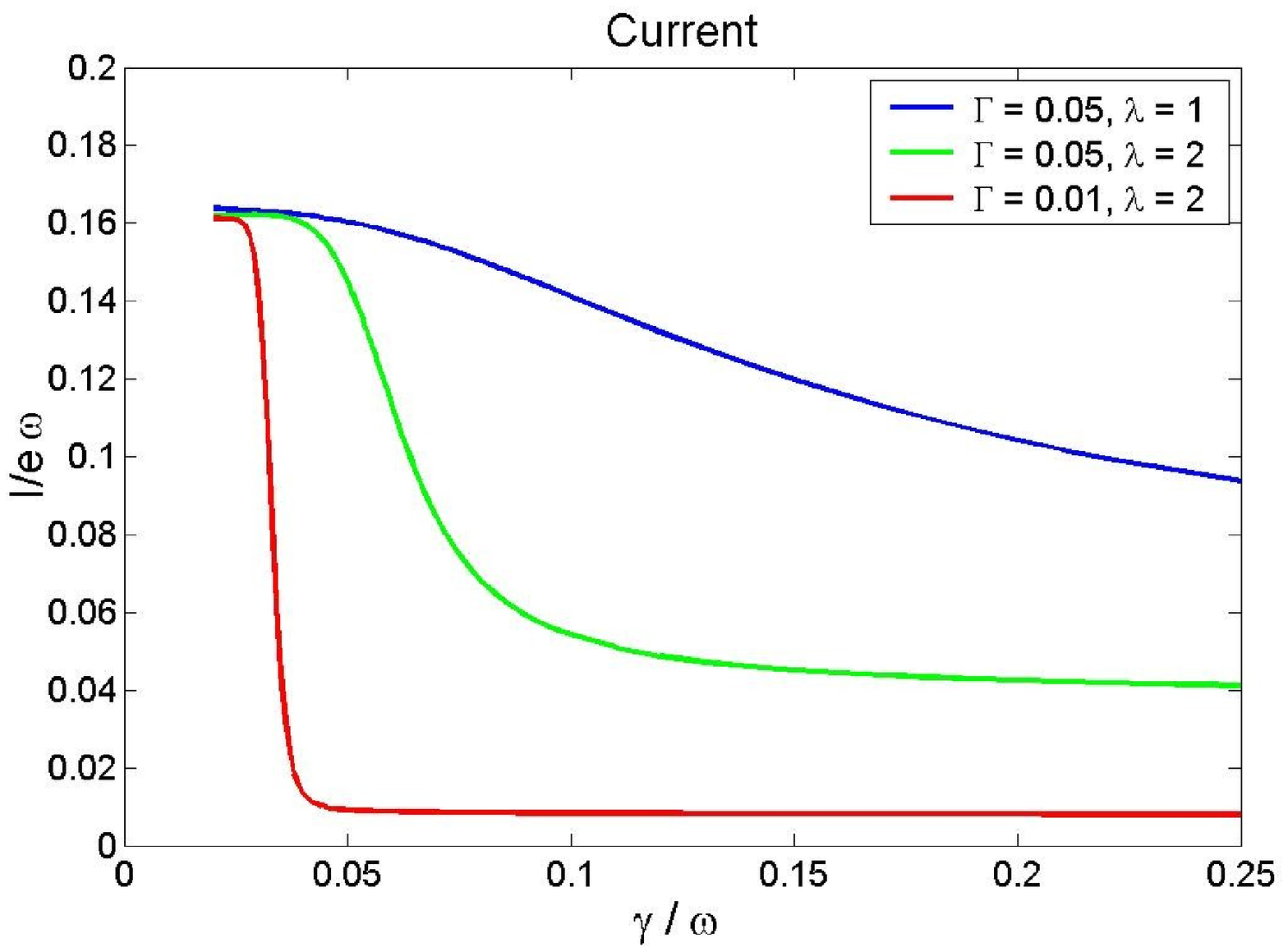}
  \caption{Stationary current as a function of damping rate.}
  \label{fig3}
\end{IWCEfigure}
More dramatic effects are observed in the Fano factor, given in
Fig. 4.  Again, we just list the essential features: (i) For high
values of damping, the noise has its Poissonian value ($\simeq
1$); (ii) At low values of damping, the noise is very low
(reflecting the orderly nature of the shuttling regime); (iii) At
the shuttling cross-over there occurs a large enhancement of the
noise.

\begin{IWCEfigure}
  \centering
  \includegraphics*[width=0.8\textwidth]{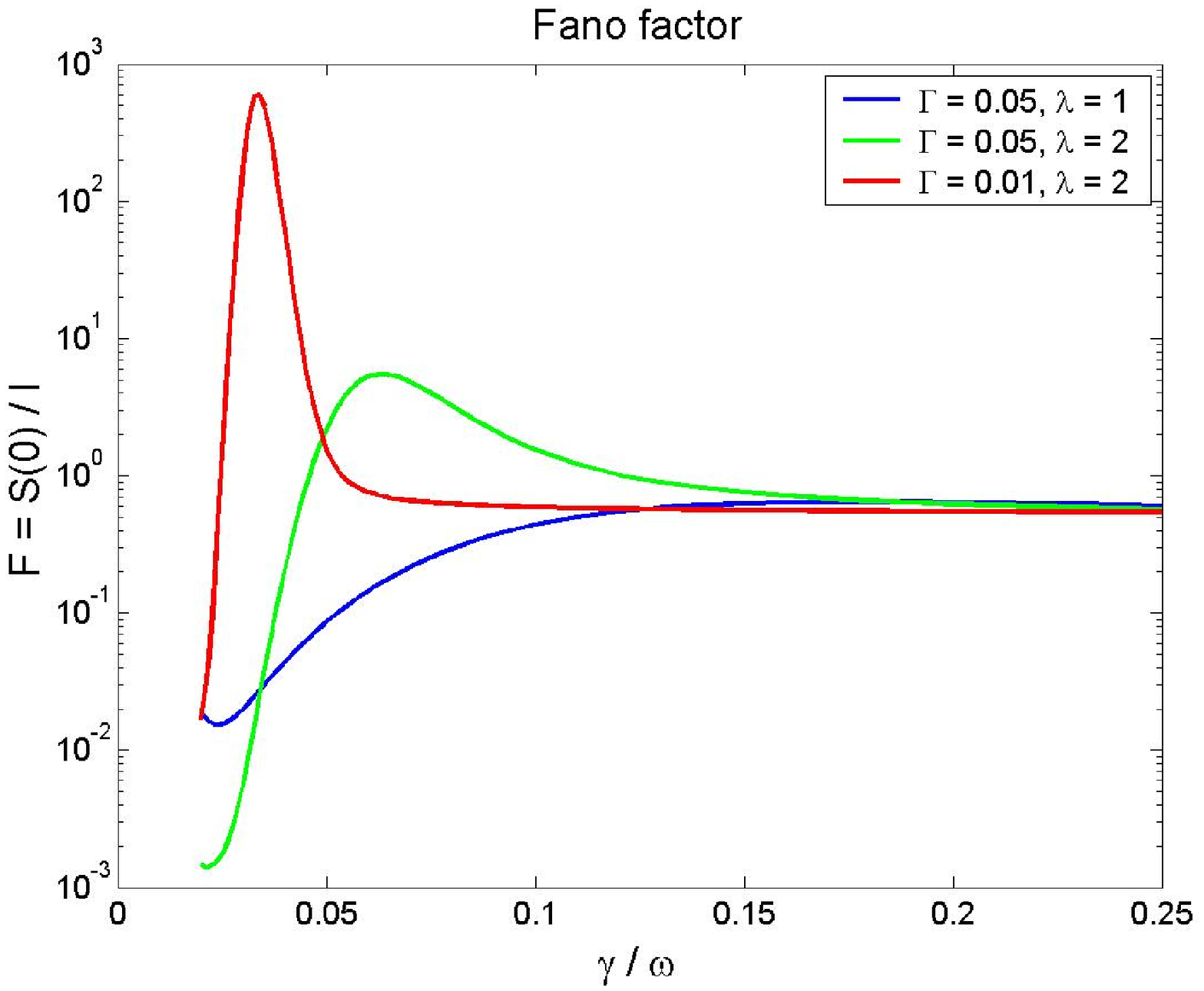}
  \caption{The Fano factor as a function of the damping.}
  \label{fig4}
\end{IWCEfigure}
The extraordinarily large values of the Fano factor of the order
of 600 can be explained  as being a consequence of a slow
switching process between two competing current channels
(shuttling and tunnelling), and in Ref. \cite{fli-epl-04} we give
a detailed analysis of this phenomenon, also supported by
semianalytic considerations.

In conclusion, we have presented a numerical technique for solving
the generalized Master equation governing a generic quantum
nanoelectromechanical device. The obtained numerical results are
interpreted with the help of phase space representations. We
believe that the methods discussed here are also applicable to
many other quantum transport situations, where the matrix
representations of the relevant operators are very large, but
where only certain extremal eigenvalues are important.

\end{multicols}

\begin{thebibliography}{99}
\bibitem{remark}
We show in  \cite{nov-prl-03,nov-prl-04,fli-prb-04} that the
off-diagonal elements $\rho_{01/10}$ decouple from $\rho_{00/11}$,
and are not needed in the evaluation of the stationary
limit.
\end{thebibliography}
\end{document}